\documentclass[prb,showpacs,superscriptaddress,footinbib]{revtex4}

\usepackage{epsfig}
\usepackage{dcolumn}
\usepackage{bm}
\usepackage{amsmath}
\usepackage{amsfonts}
\usepackage{amssymb}
\usepackage{graphicx}%

\begin{document}

\newcommand{\sign}{\operatorname{sign}}
\newcommand{\Ci}{\operatorname{Ci}}
\newcommand{\Si}{\operatorname{Si}}
\newcommand{\tr}{\operatorname{tr}}

\newcommand{\beq}{\begin{equation}}
\newcommand{\eeq}{\end{equation}}
\newcommand{\beqn}{\begin{eqnarray}}
\newcommand{\eeqn}{\end{eqnarray}}

\newcommand{\slp}{\raise.15ex\hbox{$/$}\kern-.57em\hbox{$ \partial $}}
\newcommand{\lnA}{\raise.15ex\hbox{$/$}\kern-.57em\hbox{$A$}}
\newcommand{\unmedio}{{\scriptstyle\frac{1}{2}}}
\newcommand{\uncuarto}{{\scriptstyle\frac{1}{4}}}

\newcommand{\trial}{_{\text{trial}}}
\newcommand{\true}{_{\text{true}}}
\newcommand{\const}{\text{const}}

\newcommand{\intp}{\int\frac{d^2p}{(2\pi)^2}\,}
\newcommand{\intx}{\int_C d^2x\,}
\newcommand{\inty}{\int_C d^2y\,}
\newcommand{\intxy}{\int_C d^2x\,d^2y\,}

\newcommand{\bP}{\bar{\Psi}}
\newcommand{\bc}{\bar{\chi}}
\newcommand{\hs}{\hspace*{0.6cm}}

\newcommand{\bra}{\left\langle}
\newcommand{\ket}{\right\rangle}
\newcommand{\bracket}{\left\langle\,\right\rangle}

\newcommand{\D}{\mbox{$\mathcal{D}$}}
\newcommand{\N}{\mbox{$\mathcal{N}$}}
\newcommand{\Lag}{\mbox{$\mathcal{L}$}}
\newcommand{\V}{\mbox{$\mathcal{V}$}}
\newcommand{\Z}{\mbox{$\mathcal{Z}$}}
\newcommand{\A}{\mbox{$\mathcal{A}$}}
\newcommand{\B}{\mbox{$\mathcal{B}$}}
\newcommand{\C}{\mbox{$\mathcal{C}$}}
\newcommand{\E}{\mbox{$\mathcal{E}$}}


\title{Pumping current and conductance of a Luttinger liquid in the presence of two
time-dependent impurities at finite temperature}
\author{Mariano J. Salvay}
\affiliation{Departamento de F\'{\i}sica, Facultad de Ciencias
Exactas, Universidad Nacional de La Plata and IFLP-CONICET, CC 67,
 1900 La Plata, Argentina.}

\begin{abstract}We study the pumping current and the conductance in a Tomonaga-Luttinger
liquid in the presence of two time-dependent point like weak
impurities, taking into account finite-temperature effects. We
investigate the different regimes which can be established as function
of the frequency, the temperature
and the separation between the impurity potentials.  We show how
the previous zero temperature or single impurity results
are distorted.
\end{abstract}
 \maketitle

In recent years there has been an intense focus on the analysis of non-equilibrium situations in the context of electrons in low dimensionality \cite{Chamon1}.  In particular, the problem of electronic transport through a time-dependent perturbation has been studied in relation to the X-ray excitation \cite{Gogolin1} and the possibility of charge and spin exchange on conductors and semiconductors \cite{Chamon2}.  The investigation of the role of dynamic sources in highly correlated electron systems in 1D reveals an interesting equivalence with quantum evaporation of helium superfluids experiments \cite{Nature}.  It shows that a phonon source, which represents  a time-dependent perturbation, embedded in superfluids acts exciting particles so that these acquire an energy greater than the one necessary to escape from the condensate.  Possible experimental realization are a pump laser applied on a carbon nanotube producing a periodic deformation in the network structure that can be understood as an effective time-dependent impurity \cite{Science}.  If the electronic transport through the nanotube changes significantly in the presence of the perturbation, that may be used to gain information on the causes of the oscillation for the application of nanotubes as sensors or detectors \cite{Science2}.  Another possible experimental  realization is a Hall bar with a constriction \cite{Miliken}.

In the study of dynamic impurities in Luttinger liquids, two observables of special interest are the dc component of the backscattered current $I_{bs}$ and the correction to the differential conductance $\Delta G$.  For a point like time-dependent oscillatory impurity, the conductance of a one-channel quantum wire is greater than its background value $e^{2}/h$ for strong repulsive interaction (Luttinger liquid parameter $K < 1/2$) \cite{Feldman}.  This result was obtained at zero temperature.  Later, the effect of the finite length of the wire and the finite temperature on $I_{bs}$ and in the shot noise $S$ were analyzed \cite{Dolcini} \cite{Cheng}.  In another direction, some authors have considered the role of extended impurities (like rectangular barriers) in the conductance of Fermi and Luttinger liquids \cite{yo 1} \cite{makogon2}.

More recently, the effect of several time-dependent impurities was considered at zero temperature and infinite length.  For the case of two impurities oscillating with the same frequency, the dc component of $I_{bs}$ is positive even for weak repulsive interactions, due to the presence of the interference term induced by spatial correlations \cite{makogon}. The pumping current, i.e. the persistence of a dc current even in the absence of external voltage, was studied in \cite{theories}, where a power-law dependence with the frequency with an exponent $ 2K - 1$  was found.  These authors also show that this current is  proportional to the sine of the phase difference and the sine of the separation between barriers.

In this work, we study the transport properties in a Tomonoga-Luttinger liquid in presence of two point like time-dependent impurities, both oscillating with the same frequency and amplitude.  We will consider the effect of finite temperature and thus expand the results obtained in Ref. 13  and 14 at zero temperature.  By performing a perturbative expansion in the backscattering amplitude and using the Keldysh technique \cite{Keldysh}, we obtain an analytical expression for $I_{bs}$. We focus our attention on the value of $I_{bs}$ at zero external voltage, showing how it changes in relation to the zero temperature case. From  $I_{bs}$ we compute and analyze  $\Delta G$.  These quantities are presented as functions of $K$ and they are studied in two scale regimes: one that relates the temperature and the frequency and other that combines the frequency with the spatial separation between impurities.

As the computational starting point, let us consider the following
Lagrangian density, which is derived using the usual bosonization
technique:\begin{equation} L = L_{0} + L_{imp} \,  , \end{equation} where\begin{equation} L_{0} = \frac{1}{2}\Phi(x,t)
(v^{2}\frac{\partial^{2}}{\partial_{x}^{2}} -
\frac{\partial^{2}}{\partial_{t}^{2}})\Phi(x,t) \, , \end{equation}  describes
a spinless Tomonaga-Luttinger liquid with renormalized velocity  $v$ and\begin{equation} L_{imp}= - \frac{
g_{B}}{\pi \hbar \Lambda} \sum_{\pm} \delta(x - x_{\pm}) \cos[\Omega
t + \delta_{\pm}]\cos[2 k_{F}x/\hbar + 2\sqrt{\pi K
v}\Phi(x, t) + e V t/\hbar] \, ,\end{equation}represents the interaction
of spinless electrons whit two dynamical impurities located at the
points $x_{+}$ and $x_{-}$, with initial phases $\delta_{+}$ and
$\delta_{-}$ and both oscillating with frequency $\Omega$ and coupling
amplitude $g_{b}$. $V$ is the external voltage applied to the quantum wire and $K$
measures the strength of the electron-electron interactions.  For
repulsive interactions $K < 1$, and for noninteracting electrons $K
= 1$. $\Lambda$ is a short-distance cutoff. In the above expression we only take into account  backscattering between electrons and impurities, because the forward scattering does not change the transport properties studied here, at least, in the lowest-order of the perturbative expansion in the couplings.

In the absence of the impurities, the background current is $I_{0} = e^{2} V/h$ .  In the presence of the impurities the total current is $I = I_{0} - I_{bs}$.  The operator associated to the backscattered current is defined as \cite{makogon}\begin{equation}\widehat{I}_{bs}(t) = \frac{g_{B} e}{\pi \hbar \Lambda} \sum_{\pm}  \cos[\Omega t
+ \delta_{\pm}]  \sin[2 k_{F}x_{\pm}/\hbar + 2\sqrt{\pi K
v}\widehat{\Phi}(x_{\pm}, t) + e V t/\hbar] \, .\end{equation}

The backscattered current at any time t is given by\beq  I_{bs}(t)  = \langle 0|S(- \infty ;
t)\widehat{I}_{bs}(t)S(t ; - \infty) | 0 \rangle \, ,  \label{uno} \eeq where $ \langle 0|$
denote the initial state and $S$ is the scattering matrix, which to the lowest order in the coupling $g_{B}$ is given by\beq S(t ; - \infty) = 1 - i \int^{\infty}_{-\infty} d x \int^{t}_{-
\infty} L_{imp} (t') d t' \, .\eeq

In order to compute (\ref{uno}) we have first derived expression for the v.e.v. of exponentials of the $\Phi$-fields at finite temperature:\begin{multline} \langle 0|\exp[i 2 \sqrt{\pi K v} \widehat{\Phi}(x', t')]\exp[- i 2 \sqrt{\pi K v} \widehat{\Phi}(x, t)] 0 \rangle    -  \langle 0|\exp[- i 2 \sqrt{\pi K v} \widehat{\Phi}(x, t)]\exp[i 2 \sqrt{\pi K v} \widehat{\Phi}(x', t')] 0 \rangle  \\ =  \frac{ (\Lambda \pi k_{b} T / v)^{2 K} 2 i \sin[\pi K] \Theta(v | t - t'| - |x - x'|)}{|\sinh[( \pi k_{b} T / v)(x - x' - v (t - t'))] \sinh[( \pi k_{b} T / v)(x - x' + v (t - t'))]|^{K}} \,  , \label{tres} \end{multline}where $\Theta$ is the Step function.

In realistic systems the frequency $\Omega$ is expected to be quite high so that it is unlikely  that the explicit time resolution of $I_{bs}(t)$ would be experimentally accessible.  Then, it is natural to consider the time average over the period of the impurities, which can be identified  with the dc component of the backscattered current:\beq I_{bs} = \frac{\Omega}{2 \pi} \int_{0}^{\frac{2 \pi}{\Omega}} d t  I_{bs}(t)  \label{dos} .\eeq
\begin{figure}\begin{center}
\includegraphics{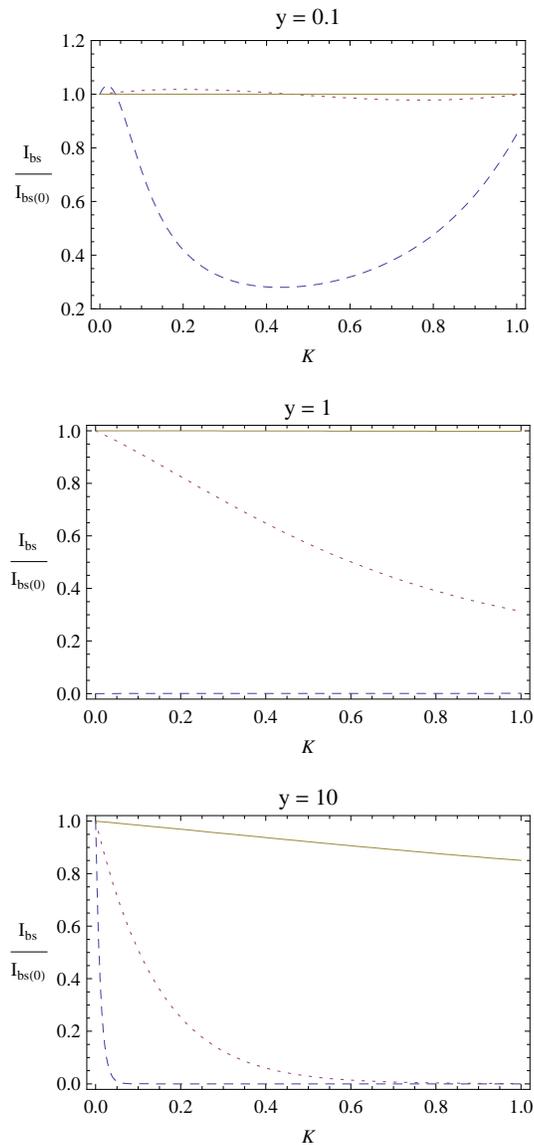}
\caption{\label{fig1ne}: Pumping current at finite temperature divided by the pumping current at zero temperature as function of $K$ and for different values of $y$. Dashed line corresponds to $z = 0.1$, dotted line to $z = 1$ and solid line to $z = 10$.}
\end{center}\end{figure}

Using (\ref{tres}) in the computation of (\ref{uno}) an inserting this in (\ref{dos}) we can compute $I_{bs}$.  Defining the dimensionless scaling parameters: $z_{\pm} = \frac{\hbar (e V / \hbar \pm \Omega)}{2 \pi k_{b} T}$, $z = \frac{\hbar \Omega}{2 \pi k_{b} T}$, $y_{\pm} = \frac{a (e V / \hbar \pm \Omega)}{v}$ and $y = \frac{a \Omega}{v}$, $I_{bs}$ can be expressed as:\begin{multline} I_{bs} = \frac{e g_{B}^{2} \Omega \sin [\pi K]
}{4 \pi^{2} \hbar^{2} v^{2}} (\frac{\Lambda \Omega}{v})^{2 K -
2}z^{1 - 2 K}   \{ ( - i \cos[\frac{2 k_{F} a}{\hbar} + \phi]\exp[ - K y / z]\Gamma[1 - K]\exp[i y_{+}] \\ \times \frac{\Gamma[ K - i z_{+}]}{\Gamma[ 1- i z_{+}]} F(K,K - i z_{+}, 1 - i z_{+}, \exp[- 2 y / z])  - i \frac{\Gamma[1 - 2 K] \Gamma[K - i z_{+}]}{ \Gamma[1 - K - i z_{+}]} + \ast )
+ z_{+} \rightarrow z_{-} \, \, , y_{+} \rightarrow y_{-} \, \,
and \, \, \phi \rightarrow - \phi \}\, .
\label{cinco} \end{multline}

In this expression $ a = x_{+} - x_{-} $ represents the spatial separation between the two impurities and
$\phi = \delta_{+} - \delta_{-}$ the phase difference.  $\Gamma$ is the Gamma function and $F$ is the Gauss hypergeometric function $_{2}F_{1}$.  Thus, we have obtained an analytical expression for the backscattered current as function of all the parameters of the system, at the lowest-order in the impurity coupling $g_{b}$.  The variables $z$ and $y$ characterize the scale regimes of the system: $z \gg 1 (\ll 1)$ is the  low (high)-temperature regime, for fixed frequency; and  $y \gg 1 (\ll 1)$ is the high (low)-frequency, with respect to the spatial separation.

We first focus our analysis to the case of pure pumping, $V = 0$. The backscattered pumping current is:\begin{equation} I_{bs} = \frac{e g_{B}^{2} \Omega}{2 \pi \hbar^{2} v^{2}} (\frac{\Lambda \Omega}{v})^{2 K - 2}\sin[\frac{2 k_{F}
a}{\hbar}] \sin[\phi]\exp[ - K y / z] z^{1 - 2 K}\{ \frac{i
 \Gamma[K - i z] \exp[i y] F(K,K - i z, 1 - i z, \exp[- 2 y / z])}{\Gamma[K] \Gamma[1 - i z]} +
 \ast \}. \label{pu}\end{equation}

This expression is the generalization for finite temperatures of the result obtained in Ref. 14.  The factors $\sin[\frac{2 k_{F}
a}{\hbar}]$ and $\sin[\phi]$ are characteristic of a pumping current in one-dimensional systems, and show that the direction of $I_{bs}$ at zero voltage is determined by the spatial separation between impurities and by the phase difference between them.  In the scale regime of low temperatures ($z \gg 1$), the pumping current goes as $\Omega^{2 K  - 1}$ for small frequency  ($y \ll 1$); and goes as $a ^{-K} \Omega^{ K  - 1} $ for high frequency ($y \gg 1$).

We note that for low temperatures and $K < 1/2$, the pumping current becomes large when $\Omega$ decreases. Hence, the perturbative expansion in powers of $g_{B}$ breaks down when $\Omega \rightarrow 0$.  Using a scaling analysis we can estimate that this expansion is valid when $\frac{g_{B}}{\hbar v} (\frac{\Lambda \Omega}{v})^{K - 1} \ll 1$.  We remark that the expression (\ref{pu}) does not include the case $\Omega = 0$, where the pumping current is zero too.  All these statements imply that the current must be a nonmonotonic function of $\Omega$.  In order to determine this function one has to go beyond the lowest-order perturbative results of this work \cite{theories}.

For high temperatures ($z \ll 1$) the asymptotic behavior of the pumping current depends to the value of $y/z = \frac{2 \pi k_{b} T a}{\hbar v}$.  For $y/z \ll 1$ $I_{bs}$ goes as $\Omega T^{2 K  - 2}$.  For $y/z \simeq 1$  the pumping current is given by the sum of two terms competing with each other: one proportional to $\Omega T^{2 K  - 2}$ and other proportional to $a \Omega   T^{2 K  - 1}$.  Finally, for $y/z \gg 1$  $I_{bs}$ goes as $\sin [\frac{a \Omega }{v}] T^{2 K  - 1} \exp[-2 \pi K k_{b} T a/\hbar v]$ .  We observe that the collapse at $\Omega \approx 0$ disappears and then the pumping current goes to zero when $\Omega$ decreases.

Figure \ref{fig1ne} shows the ratio between the pumping current at finite temperature and zero temperature.  For low frequency, $I_{bs}$ changes in relation to the case of zero temperature for $z \ll 1$, decreases for all $K$ except for very small $K$  (where  $I_{bs}/I_{bs}(T = 0) > 1$), while for $z \simeq 1$, $I_{bs}$  grows(decreases) slightly for high(small) interactions.   For intermediate frequency, occurs a suppression of $I_{bs}$ for high temperature and a monotonic decrease with $K$ for intermediate temperature.  Finally, for high frequency, the decrease of the backscattered current is more pronounced and tends to occur even for low temperatures; in any case, $I_{bs}/I_{bs}(T = 0)$ is a monotonically decreasing  function of $K$.  In general, for $K \rightarrow 0$, $I_{bs}/I_{bs}(T = 0) \rightarrow 1$; this is because for high interactions the effect of the impurities and temperature is irrelevant in any regime.
\begin{figure}\begin{center}
\includegraphics{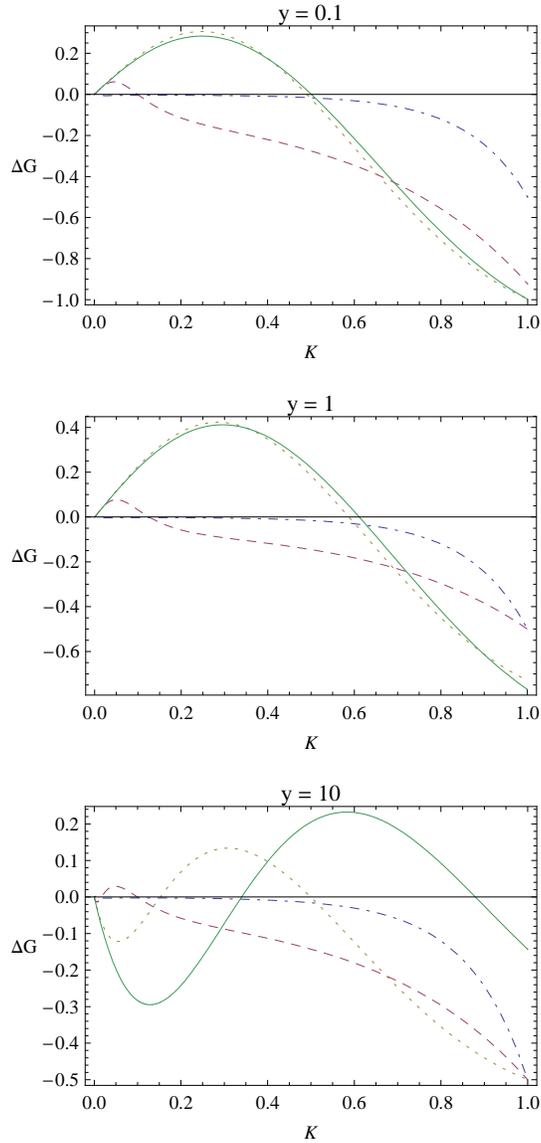}
\caption{\label{fig2ne}: Correction to the differential conductance $\Delta G$  in function of $K$ and for different values of $y$. Dotdashed line correspond to $z = 0.01$, dashed line to $z = 0.1$, dotted line to $z = 1$ and solid line to $z = 10$.  The unit is equal to $\frac{e^{2} g_{B}^{2}(\Lambda \Omega)^{2 K - 2}}{2 \pi \hbar^{3} v^{2 K}}$.  We have taken $\frac{2 k_{F}
a}{\hbar} \equiv \phi \equiv 2 n \pi$, with n integer.}\end{center}\end{figure}

From the expression (\ref{cinco}) we can obtain the correction of  the differential conductance $\Delta G =
-\frac{\partial I_{bs}}{\partial V}|_{V = 0} $ to second order in
the coupling:\begin{equation} \Delta G = \frac{e^{2} g_{B}^{2}(\Lambda \Omega)^{2 K - 2}}{2 \pi \hbar^{3} v^{2 K}} \frac{A(z, y=0, K) + \cos[\frac{2 k_{F}
a}{\hbar}] \cos[\phi] A(z, y, K)}{2} , \label{diez}\end{equation}where we have defined
\begin{multline} A(z,y,K) = \frac{z^{2 - 2 K} \exp[(i z - K) y / z] \Gamma[K - i z]}{\Gamma[K]\Gamma[1 - i z]} \{ F(K,K -i z, 1 - i z, \exp[- 2
y / z])(\Psi(K - i z) - \Psi(1 - i z) - y / z) \\ + F^{(0,1,0,0)}(K,K -i z,1
- i z, \exp[- 2 y / z]) + F^{(0,0,1,0)}(K,K - i z, 1 - i z, \exp[- 2 y / z])
+ \ast \}\, ,\end{multline}here $\Psi$ is the Digamma function, $F^{(0,1,0,0)}$ and $F^{(0,0,1,0)}$ represent the differentiation of the function $F$ respect to the second and third argument respectively.

Figure \ref{fig2ne} shows the change in the conductance of the system as function of $K$, for different scale regimes.  In the case of high temperature, the behavior is independent of the frequency regime; the effect of positive $\Delta G$ only remains in a small region of $K$ next to zero, and then corresponding at very high interaction electron-electron.  For very high temperatures this effect tends to disappear and $\Delta G$ is always negative and goes to zero for strong interactions.

For intermediate and low temperatures ($z \approx 1 $ and $z \gg 1$ respectively) the behavior of the conductance depends on the frequency regime.  In the small frequency regime the change in the conductance is almost  the same as for  a single barrier at zero temperature; this is because our definition of this regime is similar to $a \rightarrow 0$.  For intermediate frequency ($z \approx 1 $) the conductance of the system increases too for small interactions, this is for $K > 1/2$, and its value is bigger than the case of small frequencies.  For high frequency, occurs an oscillatory behavior of $\Delta G$ as function of $K$: the conductance tends to decrease for high and weak interactions, and increases in intermediate interactions.  The specific values of $K$, when $\Delta G$ changes sign, varies with $y$ and $z$.

We stress that the appearance of a pumping current and growth of the conductance even for weak electron-electron interaction ($K > 1/2$) have their origin in the spatial separation of the oscillatory impurities.  Thermal fluctuations are expected to induce decoherence, and then, at finite temperature both effects decrease in quantity.  For $z \rightarrow 0$, $I_{bs}$ is suppressed and the conductance of the system remains $e^2/h$.  Then, the effect of the impurities is irrelevant in a Luttinger liquid at very high temperature.  The exception is when $K = 1$ (Fermi liquid) and $ \frac{2 \pi k_{b} T a}{\hbar v} \leq 1$; in this case the pumping current is $I_{bs} \approx (\frac{2 \pi k_{b} T a}{\hbar v}/\sinh[\frac{2 \pi k_{b} T a}{\hbar v}])I_{bs}(T = 0)$ and the correction to the conductance is negative (in particular it is temperature-independent for a single barrier, $a = 0$). This means that for $\frac{\hbar v}{2 \pi k_{b} a} \geq  T \gg \frac{\hbar \Omega}{2 \pi k_{b}}$ Luttinger and Fermi systems are well differentiated in their transport properties.

To summarize, we presented an exact and analytical computation of the backscattered current, the pumping current and the correction to the differential conductance for a Tomonaga-Luttinger liquid in presence of two weak oscillatory impurities at finite temperature.  We analyzed the distortion of the pumping current with respect to the zero temperature case, in different scale regimes defined by the dimensionless parameters $\frac{\hbar \Omega}{2 \pi k_{b} T}$ and $\frac{a \Omega}{v}$.  We also showed how the enhancement of the conductance for $K < 1/2$, previously predicted for a single impurity, changes due to the combined effect of temperature and spatial separation of the barriers.

\vspace{1cm}
This work was partially supported by Universidad Nacional de La
Plata (Argentina) and Consejo Nacional de Investigaciones Cient\'
ificas y T\'ecnicas, CONICET (Argentina). The author is grateful
to Carlos Na\'on for a careful reading of manuscript and useful
suggestions and discussions.

\end{document}